\documentclass[prb,reprint]{revtex4-1}
\usepackage{amsmath,amssymb}
\usepackage[final]{graphicx}
\usepackage{overpic}
\usepackage{dcolumn} 
\usepackage{fancyvrb} 

\usepackage{natbib}
\usepackage{microtype}

\usepackage[colorlinks,citecolor=red,urlcolor=blue]{hyperref}
\IfFileExists{hypernat.sty}{\usepackage{hypernat}}

\newcommand{\dbar}{{d\mkern-7mu\mathchar'26\mkern-2mu}}

\newcommand{\multich}[2]{
\left.\mathchoice
  {\left(\kern-0.48em\binom{#1}{#2}\kern-0.48em\right)}
  {\big(\kern-0.30em\binom{\smash{#1}}{\smash{#2}}\kern-0.30em\big)}
  {\left(\kern-0.30em\binom{\smash{#1}}{\smash{#2}}\kern-0.30em\right)}
  {\left(\kern-0.30em\binom{\smash{#1}}{\smash{#2}}\kern-0.30em\right)}
\right.}

\newcommand{\mybinom}[2]{\binom{#1}{#2}}

\newcommand{\extbinom}[3]{\binom{#1}{#2}^{\raisebox{-0.3em}{$\scriptstyle\!\{#3\}$}}}
\newcommand{\mytimes}{}
\newcommand{\NeuschelNotation}{and this notation follows that of Neuschel}
\newcommand{\myfrac}[2]{\frac{#1}{#2}}
\newcommand{\textbinom}[2]{\mybinom{#1}{#2}}
\newcommand{\textbinomNp}[2]{\mybinom{#1}{#2}}

\newcommand{\textbinomNpQp}[2]{\mybinom{#1}{#2}}
\newcommand{\textextbinom}[3]{\extbinom{#1}{#2}{#3}}
\newcommand{\textmultich}[2]{\multich{#1}{#2}}

\newcommand{\textmultichQp}[2]{\multich{#1}{#2}}

\providecommand{\mybinom}[2]{{}_{#1}C_{#2}}
\providecommand{\extbinom}[3]{{}_{#1}C_{#2}^{\{#3\}}}

\providecommand{\mytimes}{\cdot}
\providecommand{\NeuschelNotation}{as discussed for example by Neuschel}
\providecommand{\myfrac}[2]{#1/#2}
\providecommand{\myfracPD}[2]{#1/(#2)}
\providecommand{\myfracPAll}[2]{(#1/#2)}
\providecommand{\textbinom}[2]{\text{$#1$-choose-$#2$}}
\providecommand{\textbinomNp}[2]{\text{$(#1)$-choose-$#2$}}

\providecommand{\textbinomNpQp}[2]{\text{$(#1)$-choose-$(#2)$}}
\providecommand{\textextbinom}[3]{\text{$#1$ $\{#3\}$-choose $#2$}}
\providecommand{\textmultich}[2]{\text{$#1$-multichoose-$#2$}}

\providecommand{\textmultichQp}[2]{\text{$#1$-multichoose-$(#2)$}}

\newcommand{\ns}{p} 
\newcommand{\nsp}{p} 
\newcommand{\nsmo}{{\ns-1}} 
\newcommand{\nsmop}{{(\nsmo)}} 
\newcommand{\nsgen}{\ns} 
\newcommand{\nsexpr}[2]{#1} 

\begin{document}

\title{Exploring entropy by counting microstates of the $\nsp$-state paramagnet}

\author{Steuard Jensen}
\email{jensens@alma.edu} 
\affiliation{Department of Physics and Engineering, Alma College, Alma, MI 48801}

\date{May 12, 2023}

\begin{abstract}
Moore and Schroeder proposed an effective approach to introducing entropy and the second law through computational study of models with easily countable states at fixed energy. But such systems are rare: the only familiar examples are the Einstein solid and the two-state paramagnet, which limits the available questions for assignment or discussion. This work considers the more general $\nsp$-state paramagnet and describes the modestly more complicated counting of its microstates. An instructor can draw on this family of systems to assign a variety of new problems or open-ended projects that students can complete with the help of a spreadsheet program or analytic calculation.
\end{abstract}

\maketitle

\section{Introduction}

Entropy is perhaps the most central concept in thermal physics, but students often struggle to understand it. The classical equation defining entropy $S$ from heat $Q$ and temperature $T$, $dS \ge \myfrac{\dbar Q}{T}$, leads directly to the crucial second law of thermodynamics. Total entropy always increases as sub-systems equilibriate, with heat flowing from hot to cold, since $dS\ge \myfrac{\dbar Q}{T_c} + \myfrac{-\dbar Q}{T_h} > 0$ when $T_c < T_h$.  But this provides no intuition for what entropy is.  Introductory texts often use fuzzy, inprecise language to explain that entropy somehow measures the randomness or (worse) ``disorder'' of a system.  But how does this concept relate to the mathematical definition?

The statistical definition of entropy is much more concrete than ``disorder.''  Entropy is the log of the number of microstates, $\Omega$, corresponding to a given macrostate with total energy $U$ spread over $N$ particles: $S = k \ln \Omega(N,U)$, where $k$ is Boltzmann's constant.  Temperature is then given a quantative definition relative to entropy, $1/T = \partial S / \partial U$.  So higher entropy means higher probability, and the tendency of a system to move to a more probable state hints at the second law. But it's still not clear why this should mean that heat flows from hot to cold, nor why these probabilistic statements should lead to an absolute ``law'' for macroscopic systems.

This challenge of introducing entropy in a meaningful way was discussed by Moore and Schroeder,\cite{Moore:1997ar} who proposed that students build intuition about entropy, temperature, and the second law through computational statistical calculations with systems composed of subsystems that can exchange energy.  Through this concrete exercise, students can see that the composite system's most probable microstate (with the highest entropy)  doesn't necessarily spread energy out equally among subsystems, but that the subsystem temperatures \emph{are} likely to be equal.  Introductory texts sometimes attempt similar demonstrations, but use quite small systems whose multiplicities can be computed by hand; too small to properly convey the inevitibility of the second law.  More advanced texts consider larger systems, but traditionally rely on limits and approximations that provide little intuition for students.  Computational analysis permits study of systems large enough to clearly show irreversible behavior, but still small enough for explicit calculations. 

Moore incorporated this approach into 
his introductory textbook \textit{Six Ideas That Shaped Physics} \cite{Moore:2017ut} with the help of a custom-built application to automate the process, and Chabay and Sherwood use it in their \textit{Matter \& Interactions} textbook.\cite{Chabay:2015mi} At the intermediate to advanced undergraduate level, Schroeder's textbook \textit{Thermal Physics}\cite{Schroeder:2000tx} asks students to perform their own analysis in a spreadsheet program like Microsoft Excel.  Such tools are readily available and often familiar, avoiding the need to teach formal programming.

This paper addresses a limitation of Moore and Schroeder's approach that becomes especially apparent when teaching more advanced students: there are only two physical systems whose states are easy to count explicitly, the Einstein solid and the two-state paramagnet. 
This means that there are limited options for homework and exam questions. If one system is explained in detail to illustrate the method and then students are asked to apply their understanding  to analyze the second in homework, there is no distinct third system available to use for further practice or assessment. Also, students seeking  deeper understanding have nothing to explore for larger projects or curiosity-driven questions.

Here, we show how an expanded class of systems, the ``$\nsp$-state paramagnets,'' can be analyzed using similar state counting. At the minimum value $\ns=2$, this coincides with the two-state case considered by Moore and Schroeder, and when $\nsgen \to \infty$, it is formally equivalent to the Einstein solid model. Between those extremes, explicit state counting is more complicated. But once the multiplicities are in hand the subsequent analysis proceeds the same way, and its results can teach interesting things about the relationship between the two simplest systems.

In what follows, Section~\ref{sec:systems} reviews the Einstein solid and the two-state paramagnet. Section~\ref{sec:counting} describes the general $\nsp$-state paramagnet and suggests student activities counting microstates in a spreadsheet program like Excel.  Section~\ref{sec:conclusion} summarizes the work and its limitations.
Additionally, Appendix~\ref{app:derivations} presents derivations of the state counting formulas in the main text. Appendix~\ref{app:CodeSamples} shows computer code to generate paramagnet multiplicities. Appendix~\ref{sec:analytic} shows analytical results for the large-$N$ limit.
And finally, Appendix~\ref{app:Boltzmann} reviews the more familiar analysis of these systems using the partition function.

Related approaches to teaching entropy through state counting have been considered in the literature, and some of those might connect productively to the systems considered here. For example, Schoepf\cite{Schoepf:2002sd} presented an introduction to entropy similar to Moore and Schroeder's, but with a focus on explicit tracing of energy transfers rather than computer tools. Salagaram and Chetty present a recursive algorithm for computing entropy via state counting\cite{Salagaram:2011eu} that can enable analysis of larger and more complicated systems than the basic two, and one of their applications was the three-state paramagnet.

\section{Review of basic systems}
\label{sec:systems}

A two-state paramagnet is a system of $N$ identical spin-$\myfrac{1}{2}$ particles in a magnetic field $\vec{B} = B\hat{z}$. Each has energy $-\vec{\mu} \cdot \vec{B}$, where $\vec{\mu}$ is the magnetic moment. Each spin can be up or down, $\mu_z \equiv \pm\mu$, with energy $\mp \mu B$. For ease of comparison with the Einstein solid we shift the zero of energy by $\mu B$, making the state energies 0 and $2 \mu B$. For our purposes, the important factor is just the spacing $\epsilon$ between the states, so for this system $\epsilon = 2 \mu B$.

The paramagnet's ground state occurs when all $N$ spins are parallel to the magnetic field. If we flip the spin state of $q$ spins, the total energy of the system is $U = q\epsilon$. The integer $q$ labels the macrostate, and the number of microstates is the number of ways to choose which $q$ of the $N$ spins are flipped.    The multiplicity is thus given by the binomial coefficient, often called ``$N$ choose $q$.''  Here, we denote it as $\extbinom{N}{q}{1}$:
\begin{equation}
\label{eq:twostatemult}
\Omega_2(N,q) = \extbinom{N}{q}{1}
 = \frac{N!}{q! (N-q)!} \quad.
\end{equation}
The superscript `\{1\}'  is not part of the typical binomial notation.  It denotes the fact that each spin can be chosen only once.  No spin can be ``doubly flipped" into a higher energy state because there is only one state above the ground state: repetition is not allowed.  

Below, we will discuss the more general form $\extbinom{N}{q}{r}$ (to be read as ``$N$~$\{r\}$-choose~$q$''). This allows each of the $N$ particles to be repeatedly chosen up to $r$ times.  These are called ``extended binomial coefficients'' or ``polynomial coefficients,'' \NeuschelNotation.\cite{Neuschel:2014vb}

The total ($\hat{z}$) magnetization $M$ of the paramagnet is the sum of all $N$ magnetic dipole moments. It corresponds closely to the energy: if $q$ spins are in the higher energy $\mu_z=-\mu$ state then $N-q$ will be $+\mu$, so
\begin{equation}
M 
  = \mu N \left(1 - \frac{2q}{N}\right)
  = M_\text{max} \left(1 - \frac{2U}{U_\text{max}}\right)
\quad.
\end{equation}
The maximum magnetization $M_\text{max} = \mu N$ occurs when $q=0$, and it falls to zero as $q \to N/2$.

Meanwhile, in the Einstein solid model, we approximate a solid as a system of $N$ identical, independent harmonic oscillators (this could be $N/3$ atoms, each oscillating in three dimensions about its equilibrium position). Each oscillator's energy states are separated by equal steps $\epsilon$, where this time $\epsilon = \hbar \omega$. The total energy (above the ground state) of all the oscillators is $U = q \epsilon$. The integer $q\ge 0$ again counts the total number of ``energy units'' in the system, but now $q$ has no upper bound.

For macrostate $q$, the number of microstates is the number of ways to distribute the $q$ units of energy among the $N$ oscillators. Infinite repetition is allowed: there is no limit on the maximum energy per oscillator, so
\begin{equation}
\label{eq:EinSolidMult}
\Omega_E(N,q) = \extbinom{N}{q}{\infty} \! = \extbinom{N+q-1}{q}{1}
 = \frac{(N+q-1)!}{q! (N-1)!} \,.
\end{equation}
(This is sometimes called ``$N$ multichoose $q$''.)
This expression in terms of binomial coefficients can be deduced from a ``balls and bars'' argument: one good explanation is in Section~2.2 of Schroeder's \textit{Thermal Physics} textbook.\cite{Schroeder:2000tx}

For both the Einstein solid and the two-state paramagnet, having an explicit formula for the multiplicity as a function of energy allows us to find the entropy, which in turn allows calculations of measurable quantities like temperature and heat capacity. (A full pedagogical treatment is presented in chapters~2--3 of Schroeder's \textit{Thermal Physics} textbook.\cite{Schroeder:2000tx}) Before we discuss this process, we introduce the general $\nsp$-state paramagnet.

\section{State counting and physics for general paramagnets}
\label{sec:counting}

\subsection{Finding the multiplicity}

A $\nsp$-state paramagnet is composed of $N$ identical particles each with total angular momentum quantum number $j=\myfrac{\nsmop}{2}$. 
Each particle's magnetic dipole moment can thus take the $\ns$ values $\mu_z = -j \delta_\mu, (-j+1)\delta_\mu, \ldots, j\delta_\mu$, where $\delta_\mu$ is a constant. (If the angular momentum comes entirely from electron spins, $\delta_\mu$ is two times the Bohr magneton. For other cases it may depend on the details of the system. See Schroeder p.~234 for further discussion.\cite{Schroeder:2000tx}) 

When placed in a magnetic field $B\hat{z}$, each particle thus has $\ns$ equally spaced energy states. As before we label the energy spacing $\epsilon$ (here, $\epsilon=\delta_\mu B$) and choose the ground state energy to be zero. 
For example, each particle in a three-state paramagnet has energy states 0, $\epsilon$, and $2\epsilon$. The Einstein solid, $\nsgen=\infty$, and two state paramagnet, $\nsexpr{\ns=2}{\nsmo=1}$, are limiting cases of this more general system.

As before, we define the total system energy as $U = q \epsilon$, with $0 \le q \le \nsmop N$ an integer labeling the macrostate. We count microstates by starting in the ground state and then choosing how to distribute each unit of energy among the $N$ particles.  But now \emph{limited} repetition is allowed: we can choose each particle up to $r=\nsmo$ times.   

Unfortunately, the formulas for these polynomial coefficients are more complicated than for the limiting cases $r=1$ or $\infty$. 
One formula for the trinomial ($\ns=3$) coefficients, ``$N$ $\{2\}$-choose $q$,'' was given by Andrews:\cite{Andrews:1990tr}
\begin{align}
\label{eq:trinomialformula1}
\Omega_3(N,q) =
\extbinom{N}{q}{2} &= \sum_{j=0}^{\min(q,2N-q)}
  \!\!\!(-1)^j \mybinom{N}{j} \mytimes \mybinom{2N-2j}{q-j}
\;.
\end{align}
A similar formula for the general case was given by Dani\cite{Dani:2011se}:
\begin{align}
\nonumber
\Omega_\ns(N,q) &=
\extbinom{N}{q}{\nsmo} \\
\label{eq:extbinomformula}
 &= \sum_{j=0}^{\lfloor q/\nsp \rfloor}
  (-1)^j \mybinom{N}{j} \mytimes \mybinom{N+q-\nsp j-1}{q-\nsp j}
\quad,
\end{align}
where $\lfloor q/\nsp \rfloor$ denotes the ``floor'' of $q/\ns$, which here equals the integer part of the quotient. 
Proofs of both of these formulas are given in Appendix~\ref{app:derivations}.  They use more sophisticated combinatorics arguments than the ``balls and bars'' derivation for Eq. \ref{eq:EinSolidMult}, so instructors might not find it worth working through these details.

Instead, an instructor could simply provide students with a table giving the multiplicity as a function of $q$ (the first and second columns of Table~\ref{t:threeStateHeatCap} shows an example). Such a list can be generated directly by a computer and then exported as a data file or into a spreadsheet. Specific implementations using an Excel VBA function, Mathematica, and Python are given in Appendix~\ref{app:CodeSamples}.

For more extensive projects studying the behavior of $\nsp$-state paramagnets, students may benefit from using the formula directly. Though the full proofs are better left to specialized math courses, students can gain confidence in these formulas by explicitly enumerating states for some very small values of $N$ and checking that the results match. 
A comparison of the coefficients for several values of $\nsgen$ for the cases of $N=3$ and~6 is given in Table~\ref{t:CompareCoeffs}.  It is noteworthy that for fixed $N$ and small enough $q$, all values of $\nsgen$ reach the same or nearly the same multiplicity because the energy is dilute enough that overlap is rare.

\begin{table}\begin{center}
\caption{\label{t:CompareCoeffs}A comparison of extended binomial coefficients $\extbinom{N}{q}{\nsmo}$ for several values of $N$ and $\nsgen$. The $\ns=\infty$ case corresponds to the Einstein solid. At low temperatures (small $q$) and a given $N$, all values of $\nsgen$ give similar results.}
\begin{ruledtabular}
\begin{tabular}{cc|rcccccccccc}
$N$ & $\ns$ & $q=0$ & ~1~ & ~2~ & ~3~ & ~4~ & ~5~ & ~6~ & ~7~ & ~8~ & ~9~ \\
\hline
3   &   2  &     1  & 3 &  3 &  1 &  0 &  0 &  0 &  0 &  0 &  0\\
3   &   3  &     1  & 3 &  6 &  7 &  6 &  3 &  1 &  0 &  0 &  0\\
3   &   4  &     1  & 3 &  6 & 10 & 12 & 12 & 10 &  6 &  3 &  1\\
3   &$\infty$&   1  & 3 &  6 & 10 & 15 & 21 & 28 & 36 & 45 & 55\\
\hline
6   &   2  &     1  & 6 & 15 & 20 & 15 &  6 &  1 &  0 &  0 &  0\\
6   &   3  &     1  & 6 & 21 & 50 & 90 &126 &141 &126 & 90 & 50\\
6   &   4  &     1  & 6 & 21 & 56 &120 &216 &336 &456 &546 &580\\
6   &$\infty$&   1  & 6 & 21 & 56 &126 &252 &462 &792 &1287&2002
\end{tabular}
\end{ruledtabular}
\end{center}\end{table}

\subsection{Physics and sample problems}

\subsubsection{Individual systems}

With the ability to compute multiplicities for $\nsp$-state paramagnets now in hand, we can consider specific physics problems to assign. The most straightforward standalone problems are studies of the entropy, temperature, and heat capacity of a single $\nsp$-state paramagnet that directly parallel those discussed by Moore and Schroeder for the Einstein solid and two-state paramagnet. The instructor can choose particular values of $N$ and $\nsgen$ and provide students with a spreadsheet file giving the multiplicity for each allowed energy state as described in the previous section. For $N=50$ and $\ns=3$ an excerpt of such a list is shown in the first two columns of Table~\ref{t:threeStateHeatCap}.%
\begin{table}
\caption{\label{t:threeStateHeatCap} Entropy ($S$), temperature ($T$), heat capacity ($C$), and magnetization ($M$) for a $3$-state paramagnet with $N=50$ particles, computed in Excel based on multiplicity ($\Omega$) values calculated as described in the text. (Values in parentheses are theoretical limits, not calculated.) As with all paramagnet systems, most of the $p^N=3^{50}$ possible microstates fall in macrostates near $q=q_\text{max}/2$: see Fig.~\ref{fig:multiplicityGraphs}. Boltzmann's constant $k$ and the energy level spacing $\epsilon$ have been used to make all quantities dimensionless.}
\centering
\begin{ruledtabular}
\begin{tabular}{ccD{.}{.}{2.2}D{.}{.}{2.2}cD{.}{.}{1.2}}
$q_3$ & $\Omega_3$ & \multicolumn{1}{c}{$S_3/k$} & \multicolumn{1}{c}{$kT_3/\epsilon$} & $C_3/Nk$ & \multicolumn{1}{c}{$M/M_\text{max}$} \\
\hline
0  & 1     & \multicolumn{1}{c}{0}    & \multicolumn{1}{c}{(0)}    & {(0)} & 1.00 \\
1  & 50    & 3.91 & 0.28 & \text{---} & 0.98 \\
2  & 1275  & 7.15 & 0.33 & 0.450 & 0.96 \\
3  & 22050 & 10.00 & 0.37 & 0.526 & 0.94 \\
\vdots & \vdots & \multicolumn{1}{c}{\vdots} & \multicolumn{1}{c}{\vdots} & \vdots & \multicolumn{1}{c}{\vdots} \\
48 & $4.66\times 10^{22}$ & 52.20 & 16.9   & 0.002 & 0.04 \\
49 & $4.87\times 10^{22}$ & 52.24 & 33.8   & 0.001 & 0.02 \\
50 & $4.94\times 10^{22}$ & 52.26 & \multicolumn{1}{c}{$(\infty)$} & {(0)} & 0.00 \\
51 & $4.87\times 10^{22}$ & 52.24 & -33.8  & 0.001 & -0.02 \\
52 & $4.66\times 10^{22}$ & 52.20 & -16.9  & 0.002 & -0.04 \\
\vdots & \vdots & \multicolumn{1}{c}{\vdots} & \multicolumn{1}{c}{\vdots} & \vdots & \multicolumn{1}{c}{\vdots} \\
99 & 50    & 3.91 & -0.28 & \text{---} & -0.98 \\
100& 1     & \multicolumn{1}{c}{0}    & \multicolumn{1}{c}{(0)}    & {(0)} & -1.00
\end{tabular}
\end{ruledtabular}
\end{table}

Students can  compute the (dimensionless) entropy $S/k = \ln \Omega$ corresponding to each multiplicity, then compute temperature using $\myfrac{1}{T} = \myfrac{\partial S}{\partial U}$.  Temperature can also be made dimensionless with the substitution $U=q \epsilon$:
\begin{equation}
\frac{\epsilon}{kT}
 = \frac{\partial (S/k)}{\partial (U/\epsilon)}
 = \frac{\partial (S/k)}{\partial q}
\quad.
\end{equation}
We implement this in the spreadsheet with a ``centered difference'' form for the finite difference first derivative: the temperature for $q$ energy units can be computed as
\begin{equation}
kT_q/\epsilon  \approx \frac{\Delta q}{\Delta (S/k)}
  = \frac{2}{(S/k)_{q+1} - (S/k)_{q-1}}
\quad.
\end{equation}
(The numerator is $\Delta q = (q+1) - (q-1) = 2$.)
Similarly, heat capacity is defined as $C = \myfrac{\partial U}{\partial T}$, making the heat capacity per particle $C/N$  in dimensionless form
\begin{equation}
\frac{C}{Nk}
 = \frac{1}{N}
   \frac{\partial (U/\epsilon)}{\partial (kT/\epsilon)}
 \approx \frac{1}{N} \frac{\Delta q}{\Delta (kT/\epsilon)}
\quad.
\end{equation}
Again, this last form can be implemented in a spreadsheet.

Excerpts from the results for $N=50$ three-state particles are shown in Table~\ref{t:threeStateHeatCap}. The patterns are qualitatively similar to those for the two-state paramagnet, including the unusual negative temperature regime when $q > \nsmop N/2$. (Negative temperatures arise only in rare systems where the multiplicity sometimes goes \emph{down} as energy goes \emph{up}: usually when the system has a maximum allowable total  energy.  When these systems are near the maximum energy, they increase their entropy (ie., move to a more probable state) by giving \emph{away} energy: any negative temperature is ``hotter than infinity.'')

Just as for the two-state case, the total magnetization $M$ of the system is the sum of the contributions of each magnetic dipole moment and is thus related to the energy:
\begin{equation}
\label{eq:genMagnetization}
M = \mu \nsmop N \left( 1 - \frac{2q}{\nsmop N} \right)
  = M_\text{max} \left(1 - \frac{2U}{U_\text{max}}\right)
\,.
\end{equation}
As before, $M$ decreases from $M_\text{max} = \mu \nsmop N$ when $q=0$ to zero as $q\to \nsmop N/2$.


Students  could extend this procedure to explore patterns in the distribution of states, heat capacity, or other physical quantities for various values of $\nsgen$ or $N$. The instructor can again provide multiplicity data, or for a longer project, students might benefit from computing those values themselves.

\begin{figure}
\includegraphics[width=\linewidth]{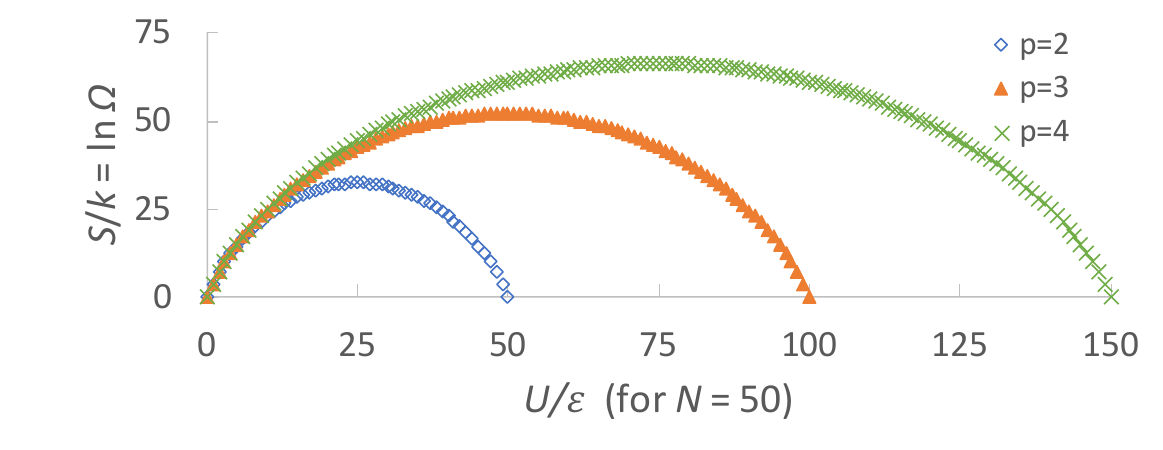}\\
\includegraphics[width=\linewidth]{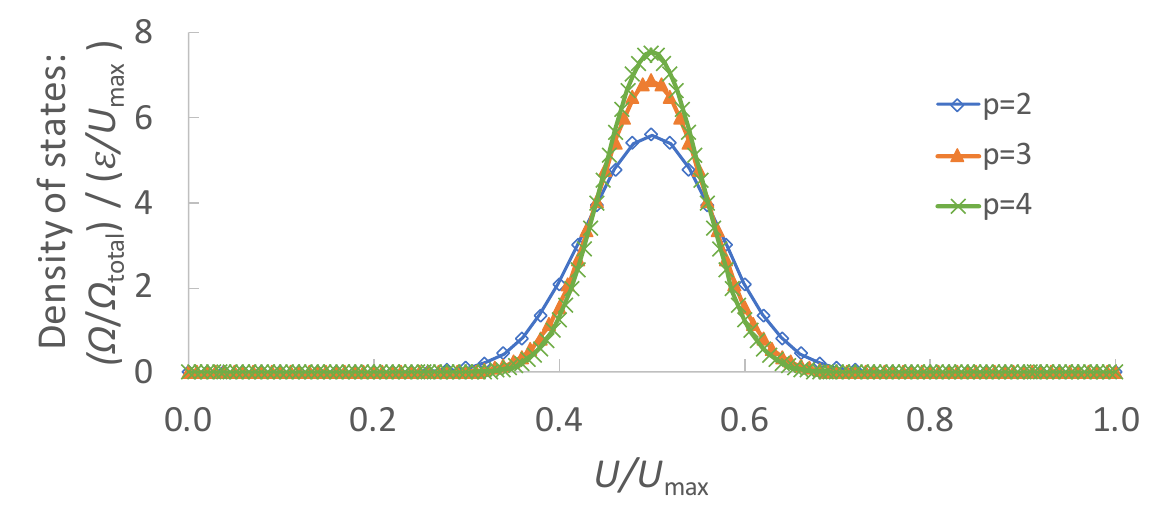}
\caption{\label{fig:multiplicityGraphs}Two possible visualizations of the distribution of multiplicities for various $\nsp$-state paramagnets (that a student could construct in Excel, using multiplicity data computed elsewhere).
Some of the options are discussed in the text: the choices leading to the top graph focus on the different overall state distributions and the matching low-energy limit, while the choices leading to the bottom one focus on the sharpness and comparative shapes of the peaks.}
\end{figure}

As an example, here we compare the multiplicity as a function of energy for paramagnets with different $\nsgen$. Figure~\ref{fig:multiplicityGraphs} shows multiplicities for systems of $N=50$ particles, where each has either $\ns=2$, 3, or 4 angular momentum states. The two panels of Figure~\ref{fig:multiplicityGraphs} show very different ways of plotting the same data.
On one level, the value of these graphs is simply to see that each paramagnet's multiplicity has a peak around the center of its allowed energy range. But this is also an opportunity for students to learn about the challenges and choices involved in effective data visualization.

Students will quickly find that graphing raw multiplicities on the same linear vertical scale is not useful because the peak heights differ by so much that a single scale cannot show all of the curves. To effectively compare the different systems, they must either use a log scale for the vertical axis (effectively graphing entropy $S = k \ln \Omega$, as in the top panel of Figure~\ref{fig:multiplicityGraphs}) or normalize it in a meaningful way (as in the bottom panel). For the horizontal axis, they can choose either to plot the energy on a common scale to show the differences in range and peak position (as in the top panel), or to scale the energy relative to the system's maximum capacity which centers all of the peaks for easier comparison (as in the bottom panel).

These different representations of the same data each have strengths and weaknesses. 
The top panel of Figure~\ref{fig:multiplicityGraphs} is optimized to show differences in the peak positions and total multiplicities for different values of $\nsgen$, and it also draws attention to the equivalent behavior of all systems in the low-energy limit. But these choices greatly obscure the sharpness of the peaks.
The bottom panel is optimized instead for comparing the sharpness and shapes of the peaks for various $\nsgen$, but its vertical normalization is subtle.%
\footnote{Because of the horizontal scaling in the top panel of Figure~\ref{fig:multiplicityGraphs}, the higher-$\nsgen$ graphs have denser data points: this leads to lower multiplicity in any given data point and thus to lower peak heights, which can be visually misleading. So rather than directly graphing the relative multiplicity $\Omega/\Omega_\text{total}$ for each state, this graph shows a fractional \emph{density} of states, normalizing by the (relative) size of the energy steps for each system. This makes the graphs easy to compare by ensuring that the area under each curve equals one.}
Confronting the surprisingly large impact of these choices in visualizing the exact same underlying data can be an excellent learning experience.


As a related project, students could compare the heat capacities or magnetizations for various values of $\nsgen$. Some care is necessary in order to study the full temperature scale, because a spreadsheet like Excel is only able to handle the size of the multiplicity numbers for $q$ up to a few hundred at most (less for higher $\nsgen$).%
\footnote{It is important when using Excel to program multiplicity formulas using the \texttt{=COMBIN(N,q)} function for the binomial coefficients rather than explicitly using factorials, because that function in Excel can handle larger arguments than the factorials can: it is coded to perform the cancellation of terms between numerator and denominator before evaluating the result. Some spreadsheet programs other than Excel do not implement \texttt{=COMBIN(N,q)} carefully in this way, and are limited to smaller systems as a result.}
Students can study smaller systems (say, $N=50$--$100$) over their full range of energy $q$ to study the high-temperature and negative temperature regime.  But small systems don't work well for low temperature, since even just one or two units of energy corresponds to fairly large temperature.  To study this low-temperature regime, students can  consider larger systems (say, $N=5000$ or more), but only at small values of energy.

\begin{figure}
\includegraphics[width=\linewidth]{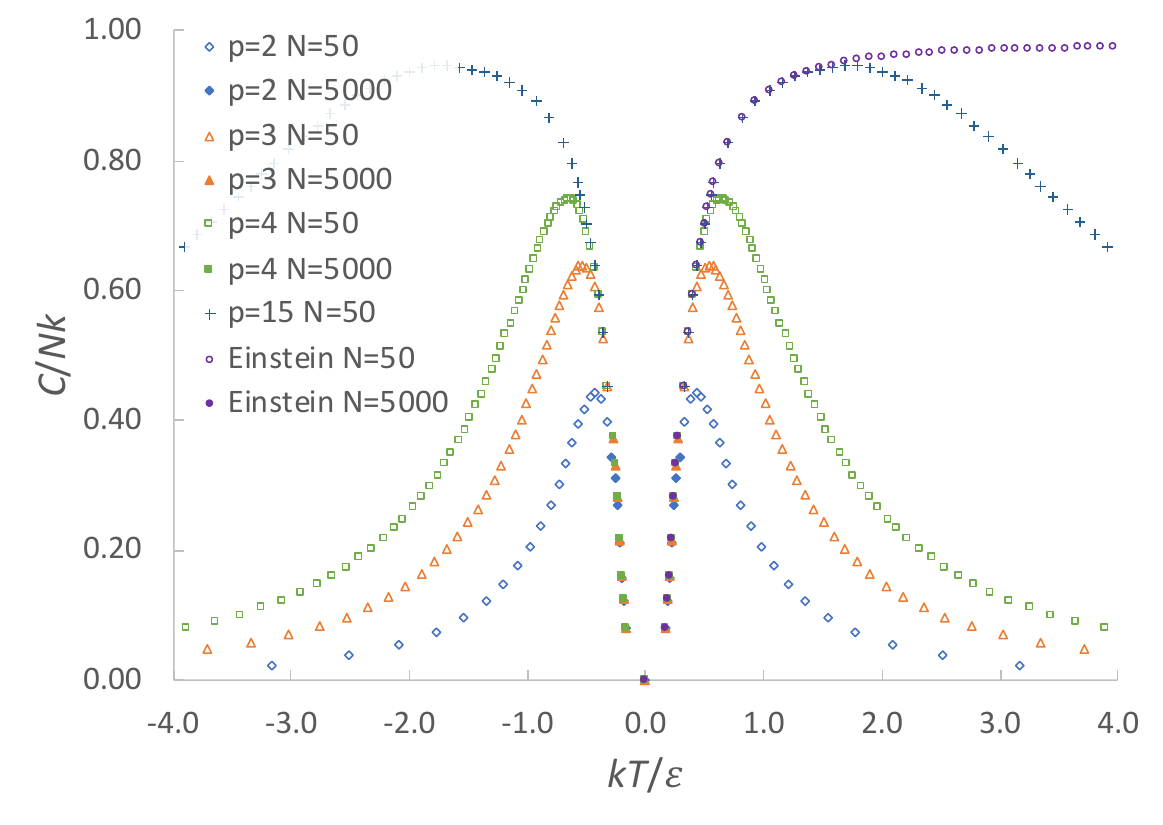}
\caption{\label{fig:comparisonHeatCaps}Heat capacities computed and graphed in Excel 
for various paramagnets and an Einstein solid. Calculations for $N=50$ particles can show high-temperature features, while low-energy calculations for $N=5000$ explore the lower-temperature limit where these systems all converge to the same behavior. For clarity, only every few data points are shown in regions where they would otherwise blur together.}
\end{figure}

Figure~\ref{fig:comparisonHeatCaps} shows what the result of such a project might look like, using systems of size $N=50$ and $5000$ to compare the heat capacities for $\ns=2, 3, 4, 15$ and the Einstein solid.
(As seen on the graph, for $N=50$ the lowest temperature with reliable data is $kT/\epsilon \sim 0.33$, corresponding to $q=2$.) 
Just as we saw for multiplicities in Table~\ref{t:CompareCoeffs}, the heat capacities at low temperature are nearly identical for all values of $\nsgen$. But at high temperature the curves interpolate between the two-state paramagnet and Einstein solid.  As $\nsgen$ increases more energy states are accessible, the behavior approaches that of an ideal harmonic oscillator, which has no limit on its energy.  Note that the Einstein solid has no regime with negative temperatures because it has no maximum in the system energy.  (For $N=5000$, the plot shows negative temperature heat capacities mirrored from the positive temperature side.) A similar project could compare graphs for magnetization.

\subsubsection{Interacting systems}

After studying entropy and temperature using a single system, students can study interacting systems to build an inutitive understanding of the second law. Here, we tune the magnetic field so the energy state spacing $\epsilon$ for a paramagnet matches that for an Einstein solid, which allows us to use the same integer  counting of energy chunks to consider interactions between the two systems. 

When we study such a combined system, we will take the total amount of energy shared between the two subsystems to be a constant, and we will distinguish the macrostates by how the energy is split between the the two.  For example, consider a system comprised of a three-state paramagnet with $N_3=50$ particles interacting with an Einstein solid with $N_E=50$ oscillators.  If the system has total energy $U_\text{total}=100\epsilon$, then the energies of the subsystems are not independent, but can vary under the constraint $q_3 + q_E = q_\text{total} = 100$.  Table~\ref{t:interactionData} presents representative data for this example,  labeled by the number of energy units $q_3$ in the paramagnet.

The critical assumption underlying statistical mechanics is that an isolated system is equally likely to be in any microstate corresponding to its macrostate.  This is true only if we consider the system over timescales that are much larger than the time the system takes to move between microstates.  Over these large timescales, all microstates will be sampled equally and we can say that the system is in equilibrium.  This system timescale is inversely proportional to the strength of the interactions: strongly interacting systems reach equilibrium faster.   

Here, we will assume that the internal interactions of each subsystem are much stronger than the interactions that exchange energy between them, meaning the subsystems reach equilibrium much faster than the combined system.  We will consider the behavior of the system at timescales larger than the subsystem internal timescales, meaning we can simply count the subsystem microstates and assume they are sampled equally.  But because we wish to understand what happens when two systems are first brought into contact and establish an equilibrium, we count the microstates of the combined system for each separate pair of values $q_3,q_E$. (Over much longer timescales, all of these pairs would be well sampled and it would only be meaningful to count the total microstates of the combined system macrostate, $q_\text{total}=100$.)

\begin{table*}
\caption{\label{t:interactionData}Multiplicities, temperatures, and probabilities for a three-state paramagnet with $N_3=50$ particles interacting with an Einstein solid with $N_E=50$ oscillators, sharing $100$ total energy units between them. Students can identify the most likely macrostate $q_3=32$ of the combined system as the one with the highest total multiplicity and greatest probability, and also the one where the individual system temperatures are most nearly equal (though $q_3=33$ is only marginally less likely).}
\centering
\begin{ruledtabular}
\begin{tabular}{ccD{.}{.}{2.2}c|cD{x}{{}\times{}}{3.4}D{.}{.}{2.2}c|D{x}{{}\times{}}{3.4}D{.}{.}{3.2}c}
$q_3$ & $\Omega_3$ & \multicolumn{1}{c}{$S_3/k$} & {$kT_3/\epsilon$}
 & $q_E$ & \multicolumn{1}{c}{$\Omega_E$} & \multicolumn{1}{c}{$S_E/k$} & {$kT_E/\epsilon$}
 & \multicolumn{1}{c}{$\Omega_\text{total}$} & \multicolumn{1}{c}{$S_\text{t}/k$} & Prob \\
\hline
0  & 1     & 0    & 0 
 & 100 & 6.71x 10^{39} & 91.70 & 2.52 & 
 6.71x 10^{39} & 91.70 & $10^{-14}\%$ \\
1  & 50    & 3.91 & 0.28
 & 99 & 4.50x 10^{39} & 91.31 & 2.50 & 
 2.25x 10^{41} & 95.22 & $10^{-12}\%$ \\
2  & 1275  & 7.15 & 0.33 
 & 98 & 3.01x 10^{39} & 90.90 & 2.48 & 
 3.84x 10^{42} & 98.05 & $10^{-11}\%$ \\
3  & 22050 & 10.00 & 0.37
 & 97 & 2.00x 10^{39} & 90.50 & 2.46 & 
 4.43x 10^{43} & 100.50 & $10^{-10}\%$ \\
\vdots & \vdots & {\vdots} & {\vdots} & \vdots
 & \multicolumn{1}{c}{\vdots} & {\vdots} & {\vdots} & \multicolumn{1}{c}{\vdots} & {\vdots} & {\vdots} \\
31 & $2.05\times 10^{20}$ & 46.77 & 1.68 
 & 69 & 4.50x 10^{33} & 77.49 & 1.87 
 & 9.23x 10^{53} & 124.26 & 7.68\% \\
32 & $3.65\times 10^{20}$ & 47.35 & 1.79 
 & 68 & 2.63x 10^{33} & 76.95 & 1.85 & 
 9.61x 10^{53} & 124.30 & 8.00\% \\
33 & $6.28\times 10^{20}$ & 47.89 & 1.90 
 & 67 & 1.53x 10^{33} & 76.41 & 1.83 & 
 9.60x 10^{53} & 124.30 & 7.99\% \\
34 & $1.04\times 10^{21}$ & 48.40 & 2.03 
 & 66 & 8.83x 10^{32} & 75.86 & 1.81 & 
 9.22x 10^{53} & 124.26 & 7.68\% 
\end{tabular}
\end{ruledtabular}
\end{table*}

Again, the instructor can  provide data for the number of microstates of the paramagnet subsystem as a function of $q_3$ so that students don't need to use Eqs.~\eqref{eq:trinomialformula1} or~\eqref{eq:extbinomformula}.   Students would be expected to construct the rest of the table using the same procedure described by Moore and Schroeder. They can find $q_E$ for each row from the constraint $q_E = q_\text{total} - q_3$, then compute the total number of microstates for the combined system for each value of $q_3$ as the product $\Omega_\text{total}(q_\text{total}, q_3) = \Omega_3(q_3) \Omega_E(q_E)$, and the net probability from the sum of the total multiplicity values over all macrostates.

Students can identify the most likely subsystem macrostate $q_3$ in several ways. The most likely configuration is the one with the highest total multiplicity $\Omega_\text{total}$, and correspondingly the highest entropy. Just as important, it is also the state where the temperatures of the two subsystems are most nearly equal. (The instructor could frame the problem to ask students to apply a specific approach, or to ask them to comment on both.) 

In this particular example, the calculation excerpts in Table~\ref{t:interactionData} make it clear that the macrostates with $q_3 = 32$ and~$33$ are the most likely, with $q_3=32$ marginally favored. Some students may at first be surprised that two systems of ``equal size'' would not share the energy equally: this gives an opportunity to emphasize the central importance of multiplicity in determining the probabilities of states rather than just energy. This reinforces Moore and Schroeder's lesson that even though all macrostates are possible, the overwhelming majority of the probability is concentrated near the maximum entropy state: the essence of the Second Law.



\begin{figure}
\centering
\includegraphics[width=\linewidth]{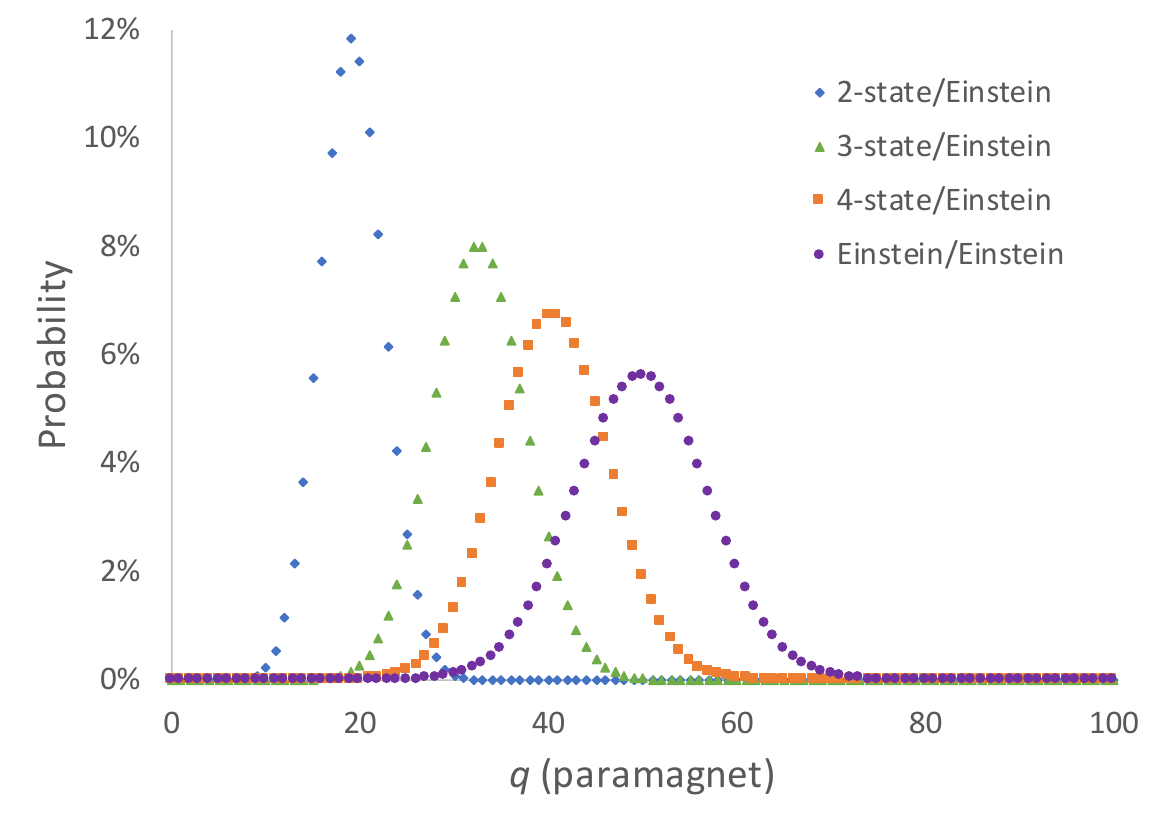}
\caption{\label{fig:comparisonProbs}Interaction macrostate probabilities for three different paramagnetic systems with $N=50$ particles with an equal-sized Einstein solid and 100 total units of energy, as well as an interaction between two Einstein solids. As $\nsgen$ increases, the particles interact more and more like a second Einstein solid.}
\end{figure}

As before, interactions involving different $\nsp$-state paramagnets provide opportunities for more open-ended exploration. An interested student could use these methods to consider interactions between an Einstein solid and  paramagnets with different values of $\nsgen$, and study the patterns that arise. As one example, Figure~\ref{fig:comparisonProbs} shows the probability distributions for various macrostates for paramagnets with $\ns = 2, 3, 4$ and also for interaction between two Einstein solids.
This graph is another clear illustration of the way in which the Einstein solid is a limiting case of the paramagnet as $\nsgen \to \infty$.

Many other avenues of exploration suggest themselves. One could investigate these systems' behavior for a variety of values of total energy, including both $q_\text{total} \ll N$ and $\gg N$. Are there recognizable trends in the equilibrium temperature as a function of $q_\text{total}$? Or, for a given Einstein solid, might a two-state paramagnet with larger $N$ have similar interaction behavior as a smaller three-state paramagnet? (Would the details depend on the  total energy?) Are there interesting patterns in interactions between paramagnets of different $\nsgen$ or $N$? There is a great deal of room for students to develop their own questions.

\section{Conclusions}
\label{sec:conclusion}

For students studying thermal physics based on Moore and Schroeder's state-counting approach to entropy and the second law, the $\nsp$-state paramagnets are a natural generalization of the Einstein solid and two-state paramagnet systems that form the foundation of that entry point to the subject. The modest increase in complexity lies entirely in the initial setup, so it is straightforward to assign problems that require no student background beyond what they have already practiced. This provides a source of additional examples to supplement the standard two for homework or exams. 
(Some analytic analysis is also possible: see Appendix~\ref{sec:analytic}.)
Because these systems form a related family, there is a great deal of room for open-ended projects and curiosity-driven exploration using these same basic tools involving explicit states counting.

It is worth explicitly commenting that state counting is not the \emph{easiest} approach to studying $\nsp$-state paramagnets.  As with so many examples, once students understand  Boltzmann factors and the partition function, those tools provide a much more direct way to determine the temperature dependence of measurables like energy and heat capacity. (This is briefly reviewed in Appendix~\ref{app:Boltzmann}.) In fact, once students have learned both approaches it can be interesting to ask them to study the same system in both ways. Seeing the agreement between these methods can give a satisfying demonstration of their relationships.

\begin{acknowledgments}
The author thanks Thorsten Neuschel for helpful clarifications of the useful range of his extended binomial coefficient approximation, and also students at Alma College 
for serving as guinea pigs for some of these ideas.
\end{acknowledgments}


\appendix

\section{Derivations of multiplicity formulas}
\label{app:derivations}

The multiplicity formulas for $\nsp$-state paramagnets given in section~\ref{sec:counting} (Eqs.~\eqref{eq:trinomialformula1} and~\eqref{eq:extbinomformula}) are based on the functions ``$N$ $\{\nsmo\}$-choose $q$'', where we can choose any given one of the $N$ elements at most $\nsmo$ times.  In this appendix, we present formal mathematical derivations of those equations.

It will be helpful in what follows to recall that binomial coefficients ``$N$ choose $q$'', with $p=2$, can be interpreted as the coefficient of $x^q$ in the expansion of the binomial $(1+x)^N$: this counts the number of ways to choose $q$ terms out of the $N$ terms of the product to contribute a power of $x$ (corresponding in the two-state paramagnet to a particle contributing one unit of energy). This relationship is expressed as the binomial expansion
\begin{equation}
\label{eq:binomialexp}
(1+x)^N = \sum_{q=0}^N \mybinom{N}{q}\, x^q \quad.
\end{equation}

Our explicit example of the three-state paramagnet has energy states 0, $\epsilon$, and $2\epsilon$, which suggests a parallel to the binomial expansion from Eq.~\eqref{eq:binomialexp}:
\begin{equation}
\label{eq:trinomialexp}
(1+x+x^2)^N = \sum_{q=0}^{2N} \extbinom{N}{q}{2}\, x^q \quad.
\end{equation}
Conceptually, as before, each of the $N$ terms on the left corresponds to one of the $N$ particles. When a given term contributes an $x$ to the product, that corresponds to a particle carrying one unit of energy, and when it contributes an $x^2$ that corresponds to two units of energy. The coefficient of $x^q$ is thus the total number of ways of distributing energy to add up to $U=q\epsilon$. This same reasoning extends directly to the more general extended binomial coefficients:
\begin{equation}
\label{eq:ExtBinomExpansion}
(1+x+x^2+\dotsb+x^\nsmo)^N =
  \sum_{q=0}^{N\nsmop} \extbinom{N}{q}{\nsmo} x^q 
\quad.
\end{equation}

The equations for the extended binomial coefficients given in section~\ref{sec:counting} can both be derived using combinatorial proofs based on the ``principle of inclusion-exclusion:'' a process of repeated overcounting and corrections. 
For $\ns=3$, Andrews\cite{Andrews:1990tr} provides two formulas, the first of which was given as Eq.~\eqref{eq:trinomialformula1} above:
\begin{align}
\label{eq:trinomialformula1a}
\extbinom{N}{q}{2} &= \sum_{j=0}^{\min(q,2N-q)}
  (-1)^j \mybinom{N}{j} \mytimes \mybinom{2N-2j}{q-j}
\quad.
\end{align}
Andrews presents these formulas without proof, calling them ``easily derived,'' which might possibly be true for those with recent practice in combinatorics. Here, it seems worth explaining in more detail.

\begin{figure}
\begin{overpic}[percent,width=0.85\linewidth]{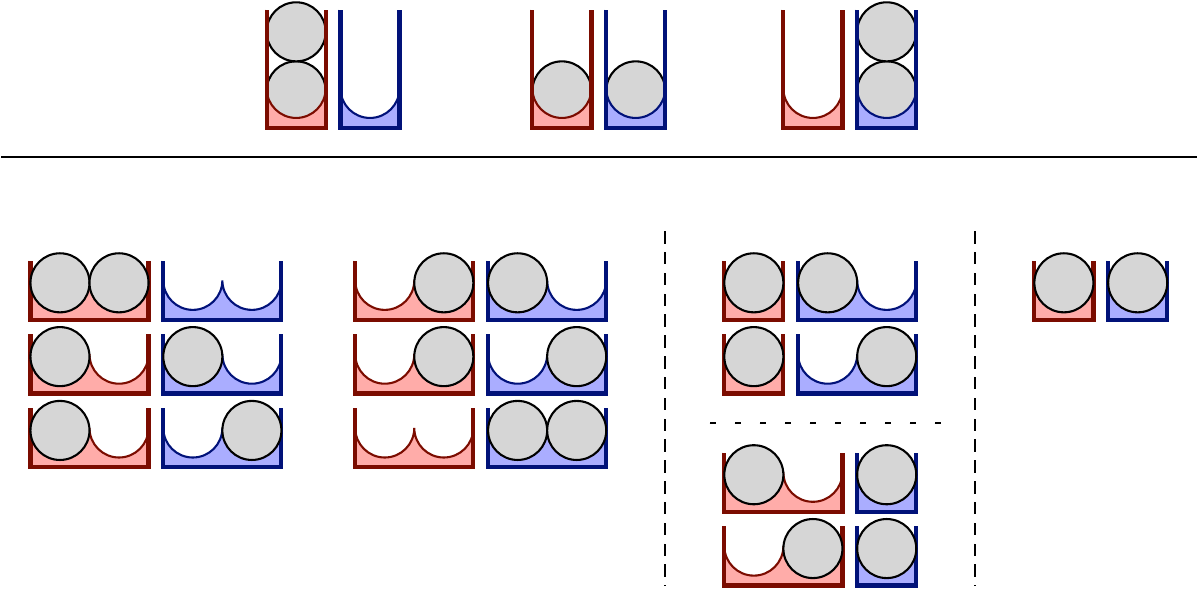}
  \put(100,43){${}={}\quad 3$}
  \put(21,31){$j=0$}
  \put(64,31){$j=1$}
  \put(87,31){$j=2$}
  \put(26,-5){$6$}
  \put(53,-5){${}-{}$}
  \put(67,-5){$4$}
  \put(79,-5){${}+{}$}
  \put(91,-5){$1$}
  \put(100,-5){${}={}\quad 3$}
\end{overpic}
\hfill
\vspace{0.75em}
\caption{\label{fig:InclusionExclusion} \emph{Top:} Direct state counting for $N=2$, $q=2$, $\nsexpr{\ns=3}{\nsmo=2}$: two boxes and two marbles. Each box can hold 0, 1, or~2 marbles (three possible states), so $\textextbinom{2}{2}{2}=3$.
\emph{Bottom:} Computing this using Eq.~\eqref{eq:trinomialformula1}/\eqref{eq:trinomialformula1a}, where each box has two ``slots.'' First ($j=0$) count all ways to distribute 2 marbles among 4 slots. This double counts cases where a box contains a single marble (since slots are equivalent), so ($j=1$) cycle through subtracting those cases: choose a box to have exactly one marble, then count ways to distribute the remaining marble. But this double subtracts cases where \emph{two} boxes have just one marble, so ($j=2$) add that back in.}
\end{figure}
As illustrated in Figure~\ref{fig:InclusionExclusion} for the case $N=2$, $q=2$, we apply the principle of inclusion-exclusion as follows. Imagine each particle as a box that can hold up to two marbles (energy units): one in a slot on the left and one on the right. There are thus $2N$ slots, and naively we can distribute the $q$ marbles in $\textbinom{2N}{q}$ ways. (This is the $j=0$ term in the sum above.) But this double counts cases where a single box has only one marble, because it doesn't matter which slot is filled. So we choose one particular box (there are $\textbinom{N}{1}$ choices), assume it has only one marble, and subtract off the number of ways of distributing the remaining $q-1$ marbles among the remaining $2(N-1)$ slots: $\textbinomNpQp{2(N-1)}{q-1}$ ways. (This subtraction is the $j=1$ term in the sum.) But now we have overcorrected, because some of those configurations had a second box with only one marble. We need to add back the count of the number of configurations with two boxes containing one marble each. There are $\textbinom{N}{2}$ choices, and for each one the remaining count is $\textbinomNpQp{2(N-2)}{q-2}$. (This is the $j=2$ term.) We repeat this alternating sum until we have considered every possible configuration: the last nonzero term is either $j=q$ (we've used all the energy) or $j=2N-q$ (we've used all the ``empty slots''), whichever comes first. 
(Try this yourself for $N=2$, $q=3$.)

Meanwhile, the general formula in Eq.~\eqref{eq:extbinomformula} is
\begin{equation}
\extbinom{N}{q}{\nsmo} = \sum_{j=0}^{\lfloor q/\nsp \rfloor}
  (-1)^j \mybinom{N}{j} \mytimes \mybinom{N+q-\nsp j-1}{q-\nsp j}
\;.
\end{equation}
This was proven by Dani by direct calculation\cite{Dani:2011se}, but a counting argument may give deeper insight.

We again apply the principle of inclusion-exclusion. Start by allowing \emph{unlimited} energy per particle and count the number of ways to distribute the $q$ units of energy among $N$ particles. Just as for the Einstein solid in Eq.~\eqref{eq:EinSolidMult}, this is the ``multichoose'' function $\textmultich{N}{q} = \textbinomNp{N+q-1}{q}$. (This is the $j=0$ term in the sum above.) For finite $\nsgen$ this obviously overcounts the number of states, because some particles may have been assigned more than $\nsmo$ units of energy. We need to subtract off the number of states where at least one particle has $\ns$ or more energy units. To do this, we first identify the overfilled particle (there are $\textbinom{N}{1}=N$ choices) and assign it a baseline of $\ns$ units of energy, and then we count the number of ways to distribute the remaining $q-\nsp$ energy units among all $N$ particles: $\textmultichQp{N}{q-\nsp} = \textbinomNpQp{N+q-\nsp-1}{q-\nsp}$. (This subtraction is the $j=1$ term in the sum.) But now we have overcorrected: we have subtracted the states where \emph{two} particles have too much energy twice. To add back these overcounted states, there are $\textbinom{N}{2}$ possible pairs, and for each possible pair we must add back the number of states in which both have at least $\ns$ units of energy: $\textmultichQp{N}{q-2\ns} = \textbinomNpQp{N+q-2\nsp-1}{q-2\nsp}$. (This positive correction is the $j=2$ term.) This alternating process continues until we have exhausted the number of energy units: until $j = \lfloor q/\ns \rfloor$ (the integer part of the quotient), proving the formula above.

\section{Code for computing multiplicities}
\label{app:CodeSamples}

In most cases, asking students to write code to implement the multiplicity formulas themselves will be outside the scope of the course, but the instructor will need some way of generating the necessary data. Sample code for several platforms is given below, and implementation files for each are included in the supplementary material.%
\footnote{Sample code can be found in online supplementary material at \protect\url{https://pubs.aip.org/ajp/article-supplement/2820263/zip/736_1_5.0061383.suppl_material/}.}

One straightforward way to generate the necessary data is to define a custom function in a spreadsheet program. The instructor could use this to generate a fixed data table to provide to students, or if all of the students are using the same program they could be given a spreadsheet template with the necessary function in place. For example, Excel on any platform can use the following VBA code to implement Eq.~\eqref{eq:extbinomformula}:\footnote{D.~Koutsoyiannis, private communication.}
\begin{verbatim}
Function ExtBinom(n, q, p)
result = 0
For j = 0 To q \ p
 result = result + (-1)^j * _
  Application.WorksheetFunction.Combin(n,j)* _
  Application.WorksheetFunction.Combin( _
   n+q-p*j-1, q-p*j)
 Next j
ExtBinom = result
End Function
\end{verbatim}
(In VBA, ending a line with ``space-underscore'' allows a formula to split across lines.
The Excel \texttt{COMBIN(N,q)} function referenced here is the ``combinations'' function that computes the binomial coefficients $\mybinom{N}{q}$.)
This custom function can then be used in spreadsheet formulas as \texttt{=ExtBinom(n,q,p)}. 

However, Excel's use of floating point math makes this unreliable for moderate to large arguments. The supplementary material includes a more careful implementation with broader validity and error handling.

Another approach that can be effective for standalone problems is to generate the multiplicity values in advance and give them to the students as a data file or spreadsheet template. 
For example, using Mathematica a list of the $\ns=3$ multiplicities for $N=50$ derived from Eq.~\eqref{eq:ExtBinomExpansion} is produced by the command 
\verb!CoefficientList[(1 + x + x^2)^50, x]!,
and a function giving the extended binomial coefficient $\extbinom{n}{q}{\nsmo}$ is
\begin{center}
\verb!extBinomial[n_, q_, !\texttt{\nsexpr{\ns}{\nsmo}}\verb!_] :=! \hspace*{\fill} \\
\hfill 
\verb!Coefficient[Sum[x^a, {a, 0, !\texttt{\nsexpr{\ns-1}{\nsmo}}\verb!}]^n, x, q]!\ .
\end{center} 
For larger values of $N$ and $\nsgen$, this method can be slow: it may be more efficient to implement the formula of Eq.~\eqref{eq:extbinomformula} instead. That is just as straightforward to do in Mathematica as it was in VBA above, or in a general purpose programming language like Python we can write: %
\begin{SaveVerbatim}{PythonExtBinom}
import scipy.special as sp
def extBinom(n,q,p):
  result = 0
  for j in range(q//p + 1):
    result += (-1)**j * sp.comb(n,j,True) \
        * sp.comb(n+q-p*j-1,q-p*j,True)
  return result
\end{SaveVerbatim}
\protect\UseVerbatim{PythonExtBinom}
(Ending a line with a backslash splits a formula across lines.)
The function \texttt{scipy.special.comb(n,j,True)} computes binomial coefficients as exact integers. %

\section{Analytic calculations at large $N$}
\label{sec:analytic}

In addition to explicit state counting, the  multiplicities for the Einstein solid and the two-state paramagnet are simple enough for students to evaluate analytically  using Stirling's approximation when $N$ and $q$ are large. This allows  explicit calculations of quantities like entropy as a function of energy or of energy as a function temperature.

For example, for the two-state paramagnet, applying the leading terms of Stirling's approximation to the natural log of Eq.~\eqref{eq:twostatemult} gives $S/k \approx N \ln N - q \ln q - (N-q) \ln (N-q)$, so the temperature is given by $1/T = \partial S/\partial U = \myfracPAll{k}{\epsilon} \partial (S/k)/\partial q$, and thus $\epsilon/kT \approx \ln [(N-q)/q]$. Solving for energy, we find $U = q\epsilon \approx N\epsilon/(1+e^{\epsilon/kT})$, or rearranging terms, $U \approx \myfracPAll{N\epsilon}{2} \bigl(1 - \tanh(\epsilon/2kT)\bigr)$. Then from Eq.~\eqref{eq:genMagnetization}, the magnetization is $M=\mu N \tanh(\epsilon/2kT)$. Similar reasoning can be applied to the Einstein solid.

This procedure is more difficult for the general $\nsp$-state paramagnet because of the sum in Eq.~\eqref{eq:extbinomformula} for the multiplicity: this sum is finite, but in systems of realistic size it could have on the order of $10^{23}$ terms.
We can make progress in particular limits, but the proofs are quite specialized and it would be unrealistic to ask physics students to understand the details. Instead, it is best just to directly cite the relevant results for students to apply.

The most interesting and accessible limit is at high temperature, as the paramagnet approaches its maximum multiplicity, $q \to \myfrac{\nsmop N}{2}$. This is the central peak seen in the density of states in Fig.~\ref{fig:multiplicityGraphs}. For large $N$ , we can use results by Neuschel\cite{Neuschel:2014vb} (building on the work of Eger\cite{Eger:2012st}) to estimate the multiplicity. Neuschel gives a full expansion of the extended binomial coefficients in orders of $1/N$.  For our purposes the leading term is sufficient:%
\footnote{Keeping the next-leading term from Neuschel's expansion is equivalent to including the $1/12N$ terms in Stirling's approximation for $N!$, as can be verified explicitly for the $\ns=2$ case.}
\begin{align}
\sqrt{\frac{\nsexpr{(\ns^2-1)}{\nsmo(\nsmo+2)} N}{12}}
 \frac{1}{\nsp^N} \extbinom{N}{q}{\nsmo}
 &= \frac{1}{\sqrt{2\pi}} e^{-x^2/2}
    + \mathcal{O}\left(\frac{1}{N^{1/2}}\right) 
\,,
\end{align}
converging uniformly with respect to all integers $q$, with
\begin{equation}
x = \sqrt{\frac{12}{\nsexpr{(\ns^2-1)}{\nsmo(\nsmo+2)} N}}
     \left(q - \frac{N \nsmop}{2}\right)
\quad.
\end{equation}
This variable $x$ has been scaled to put the Gaussian term into standard normal distribution form: the prefactor outside the parentheses is one over the standard deviation: $\sigma = \sqrt{\nsexpr{(\ns^2-1)}{\nsmo(\nsmo+2)} N/12}$. Note that $x<0$ for positive temperatures, with $T\to \infty$ as $x\to 0$.

From the right hand side of this expression, it is clear that the Gaussian term will be dominant at large $N$ as long as $|x| \lesssim 1$, but that outside of that central peak the Gaussian will fall off roughly as $e^{-N}$ and cease to be reliable as a leading term. (This is why this result only allows calculations in the high-temperature limit.) Thus, for $|x|\lesssim 1$, we can reliably solve for the multiplicity:
\begin{equation}
\label{eq:largeNmult}
\Omega_\ns(N,q) = \extbinom{N}{q}{\nsmo}
 \approx \frac{\nsp^N}{\sqrt{2\pi N}}
    \sqrt{\frac{12}{\nsexpr{\ns^2-1}{\nsmo(\nsmo+2)}}}\,
    e^{-x^2/2}
\;.
\end{equation}
This is $\nsp^N$ times a standard normal distribution, so a sum over $x$ gives $\nsp^N$ as the total number of states, as expected.

To study the physical behavior of paramagnetic systems, the next step is to compute the entropy $S = k \ln \Omega_\ns$:
\begin{align}
S_\ns/k
 &\approx N \ln \ns - \frac{x^2}{2}
   + \frac{1}{2} \left( \ln\left(\frac{6}{\pi N}\right) - \ln(\nsexpr{\ns^2-1}{\nsmo(\nsmo+2)}) \right)
\;.
\end{align}
From this we can then apply the definition of temperature, $\myfrac{1}{T} = \myfrac{\partial S}{\partial U}$, or in dimensionless form,
\begin{align}
\nonumber
\frac{\epsilon}{kT}
 &= \frac{\partial S/k}{\partial q}
 = \frac{\partial S/k}{\partial x} \frac{\partial x}{\partial q}
 \approx -x \, \sqrt{\frac{12}{\nsexpr{(\ns^2-1)}{\nsmo(\nsmo+2)} N}} \\
 &= \frac{12}{\nsexpr{(\ns^2-1)}{\nsmo(\nsmo+2)} N}
     \left(\frac{N \nsmop}{2} - q\right)
\quad.
\end{align}
Solving for energy, we find
\begin{equation}
\label{eq:highTEnergy}
q = \frac{U}{\epsilon}
 \approx N \left( \frac{\nsmo}{2} - \frac{\nsexpr{\ns^2-1}{\nsmo(\nsmo+2)}}{12}\frac{\epsilon}{kT} \right)
\quad.
\end{equation}
(This correctly matches the high temperature limit of the earlier $\ns=2$ result.)
From the energy we can calculate the heat capacity at high temperature,
\begin{equation}
C = \frac{dU}{dT}
 = N k \, \frac{\nsexpr{\ns^2-1}{\nsmo(\nsmo+2)}}{12} \left(\frac{\epsilon}{kT}\right)^2
\quad.
\end{equation}
And finally, we can use the energy to find the magnetization using Eq.~\eqref{eq:genMagnetization}:
\begin{equation}
\label{eq:highTMag}
M 
  = \frac{\mu \nsexpr{(\ns^2-1)}{\nsmo(\nsmo+2)} N}{6} 
    \frac{\epsilon}{kT}
  = M_\text{max} \frac{\nsexpr{\ns+1}{\nsmo+2}}{6} \frac{\epsilon}{kT}
\quad.
\end{equation}
This $1/T$ dependence at high temperatures is Curie's Law.

All of these steps that follow from Eq.~\eqref{eq:largeNmult} are a direct generalization of calculations that students may have already seen for the two-state paramagnet or the Einstein solid, which could make this a natural followup activity on homework or exams.

\bigskip

The low-temperature limit is more difficult to explore in this way, though limited results are possible.%
\footnote{For the $\ns=3$ case, Eq.~\protect\eqref{eq:trinomialformula1} can be rewritten in terms of a hypergeometric function whose  $N \to \infty$ limit falls into a class studied by Cvitkovi\'{c} et~al.\cite{Cvitkovic:2017ae} But the challenging calculation involved leads to exactly the same result found (far more easily) with Stirling's approximation for $\ns=2$.}  
It is also less necessary. As previously discussed, in this dilute limit when $q \ll N$ all of the $\nsp$-state paramagnets (and the Einstein solid) are equivalent to the two-state case: when there are very few energy units to go around, it is rare for any particle to carry even one of them, let alone more. In this limit, the number of available particle states makes essentially no difference at all.


\section{Analysis using the partition function}
\label{app:Boltzmann}

Paramagnetic systems with $p>2$ states are by no means new: they have been thoroughly studied in the canonical ensemble using Boltzmann factors and the partition function. 
(See, e.g., Schroeder Problem~6.11.\cite{Schroeder:2000tx})
So it is instructive to see how the results of the microcanonical analysis and direct state counting presented above compare to the (substantially simpler) canonical methods.


The partition function of a single particle is the sum of Boltzmann factors $e^{-E/kT}$ for each of its $\ns$  states:
\begin{align}
Z_\ns &= \sum_{j=0}^{\nsmo} e^{-j\epsilon/kT} = 1 + e^{-\epsilon/kT} + \dotsb + e^{-\nsmop\epsilon/kT}
\;.
\end{align}
If we define the shorthand $y = e^{-\epsilon/kT}$, we can use polynomial division (or the geometric series) to write
\begin{align}
Z_\ns &= \sum_{j=0}^{\nsmo} y^j = \frac{1-y^\ns}{1-y}
\quad,
\end{align}
an algebraic identity that holds for any $y\ne 1$ and thus any $T \ne \pm\infty$. Plugging in  $y$ and rearranging terms produces
\begin{align}
\nonumber
Z_\ns &= \frac{e^{-\nsp\epsilon/2kT}}{e^{-\epsilon/2kT}}\,
\frac{e^{\nsp\epsilon/2kT}-e^{-\nsp\epsilon/2kT}}{e^{\epsilon/2kT}-e^{-\epsilon/2kT}}\\
 &= e^{-\nsmop\epsilon/2kT} \,
    \frac{\sinh(\nsp\epsilon/2kT)}{\sinh(\epsilon/2kT)}
\quad,
\end{align}
using the definition of $\sinh$.
The initial exponential term would be absent if we had chosen the zero of energy to be the midpoint rather than the ground state.

To study limits of $T$, we define the dimensionless variable $t \equiv kT/\epsilon$, so $y=e^{-1/t}$. 
The strict high-temperature limit is $t \gg \nsexpr{\ns > 1}{\nsmo \ge 1}$, which will apply in realistic cases where $\nsgen$ is a small integer. But we may also consider the formal limit $\nsgen \gg t \gg 1$: this is the $\ns=\infty$ case corresponding to an Einstein solid. In either of these high-temperature limits, $y \approx 1 - \myfracPAll{1}{t} + \myfracPAll{1}{2t^2} - \myfracPAll{1}{6t^3}$. If additionally $t \gg \nsgen$, $y^\ns \approx 1 - \myfracPAll{\ns}{t} + \myfracPAll{\nsp^2}{2t^2} - \myfracPAll{\nsp^3}{6t^3}$, and so
\begin{align}
\nonumber
Z_{\ns,\text{high-$T$}} &\approx \frac{\frac{\ns}{t}-\frac{\nsp^2}{2t^2}+\frac{\nsp^3}{6t^3}}{\frac{1}{t}-\frac{1}{2t^2}+\frac{1}{6t^3}}\\
 &\approx \nsp \left(1 - \frac{\nsmo}{2t} + \frac{\nsmop(\nsexpr{2\ns-1}{2\nsmo+1})}{12t^2} \right)
\quad.
\end{align}
On the other hand, in the limit $\nsgen \gg t \gg 1$, it follows that $y^\ns \ll 1$ and so 
$Z_{\infty,\text{high-$T$}} \approx 1/(1/t) = t = \myfrac{kT}{\epsilon}$.


From these results, the total energy of the system can be calculated in either of two standard ways:
\begin{align}
U_\ns &= N\bar{E}_\ns = \frac{N}{Z_\ns} \sum_{j=0}^{\nsmo} j\epsilon e^{-j\epsilon/kT}
 \;, \quad \text{or}
 \end{align}
 \begin{align}
U_\ns &= -\frac{N}{Z_\ns} \frac{\partial Z_\ns}{\partial \beta}
\nonumber \\
 &= \frac{N\epsilon}{2} \left( \nsmop
   - \nsp\coth(\nsp\epsilon/2kT)
   + \coth(\epsilon/2kT) 
  \right)
\;.
\end{align}

Once again we can look at limiting temperatures. 
In the strict high-temperature limit $t\gg \nsgen$, we can use the $x \ll 1$ expansion $\coth x \approx \myfrac{1}{x} + \myfrac{x}{3}$ to show $U_{\ns,\text{high-$T$}} \approx N \epsilon \left(\myfrac{\nsmo}{2} - \myfracPD{\nsexpr{\ns^2-1}{\nsmo(\nsmo+2)}}{12t}\right)$. This precisely matches the earlier asymptotic form from Eq.~\eqref{eq:highTEnergy}. And if $\nsgen \gg t \gg 1$, the previous expansion together with $\coth(\nsp/2t) \to 1$ means that $U_{\infty,\text{high-$T$}} \approx N\epsilon t = NkT$, exactly as given by the equipartition theorem for an Einstein solid. 

From the energy, we can immediately also find the magnetization using Eq.~\eqref{eq:genMagnetization}:
\begin{align}
M_\ns &= \mu N
  \left( \nsp\coth(\nsp\epsilon/2kT)
   - \coth(\epsilon/2kT)
  \right)
\quad.
\end{align}
In the strict high-temperature limit, $M_{\ns, \text{high-$T$}} \approx \mu N \myfracPD{\nsexpr{\ns^2-1}{\nsmo(\nsmo+2)}}{6t}$, exactly matching Eq.~\eqref{eq:highTMag} and reproducing the $1/T$ behavior of Curie's Law.

\end{document}